\documentclass[a4paper]{article}

\usepackage{microtype}
\usepackage{graphicx}
\usepackage{subfigure}
\usepackage{booktabs} 
\usepackage{hyperref}
\usepackage{amsmath}
\usepackage{amssymb}
\usepackage{mathtools}
\usepackage{authblk}
\usepackage{amsthm}
\usepackage[margin=1.25in]{geometry}

\usepackage{natbib}  

\newcommand{\be}{\mathbf{e}}
\newcommand{\bx}{\mathbf{x}}

\title{Probabilistic Rule Models as Diagnostic Layers: Interpreting Structural Concept Drift in Post-Crisis Finance}

\author[1]{Dmitry Lesnik}
\author[1,2]{Tobias Sch{\"a}fer}
\affil[1]{Stratyfy Inc., New York, New York, USA}
\affil[2]{Department of Mathematics, College of Staten Island, Staten Island, NY, USA \& Physics Program, CUNY Graduate Center, NY, USA}

\date{}

\begin{document}

\maketitle

\begin{abstract}

    Machine learning models used for high-stakes predictions in domains like credit risk face critical degradation due to concept drift, requiring robust and transparent adaptation mechanisms. We propose an architecture, where a dedicated correction layer is employed to efficiently capture systematic shifts in predictive scores when a model becomes outdated. The key element of this architecture is the design of a correction layer using Probabilistic Rule Models (PRMs) based on Markov Logic Networks, which guarantees intrinsic interpretability through symbolic, auditable rules. This structure transforms the correction layer from a simple scoring mechanism into a powerful diagnostic tool capable of isolating and explaining the fundamental changes in borrower riskiness. We illustrate this diagnostic capability using Fannie Mae mortgage data, demonstrating how the interpretable rules extracted by the correction layer successfully explain the structural impact of the 2008 financial crisis on specific population segments, providing essential insights for portfolio risk management and regulatory compliance.

\end{abstract}

\section{Introduction}
\label{introduction}

Machine learning models, particularly those used for high-stakes decisions in regulated domains like credit risk, rely on the core assumption that past data can predict future outcomes. However, this stability is often undermined by concept drift—a change in the statistical properties of the target variable due to shifts in economic, political, or market environments. When a model $M_1$ becomes obsolete, continuous monitoring and full model retraining are standard practice, resulting in a new model, $M_2$. Crucially, generating $M_2$ alone fails to isolate and explain what specific systematic changes occurred in the underlying concept, a major deficiency in environments demanding regulatory auditability and policy insight.

To address this challenge, we introduce a structured architecture rooted in Transfer Learning principles. Transfer Learning is a well-established paradigm aimed at leveraging knowledge learned from a data source to facilitate efficient adaptation in a related, but different, target domain. This approach, for instance, is critically valuable for analyzing financial time series, which are inherently prone to concept drift that affects forecasting accuracy, making models quickly obsolete.

\begin{figure}[ht]
    \vskip 0.1in
    \begin{center}
        \centerline{\includegraphics[width=0.9\columnwidth]{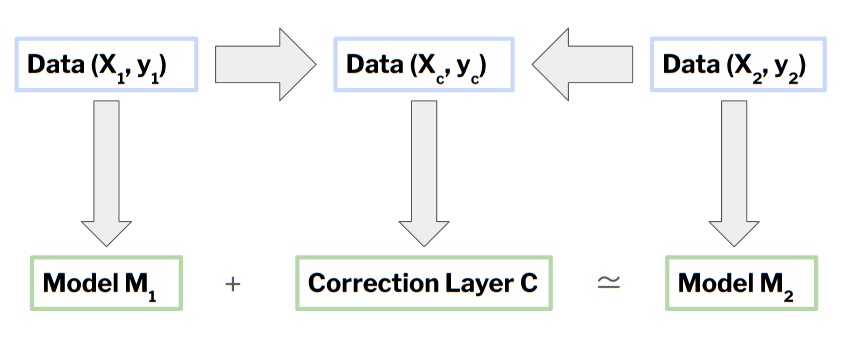}}
        \caption{Basic idea of the correction layer as an intermediate model between models $M_1$ and $M_2$.}
        \label{fig:correction_layer}
    \end{center}
    \vskip -0.2in
\end{figure}

Imagine the base model, $M_1$ was trained at an initial time $t_1$ on source data $(X_1,y_1)$ where $X_ 1$ is the input data and $y_1$ is the corresponding output. When the environment shifts and a new model $M_2$ trained on new data $(X_2,y_2)$ at time $t_2>t_1$ is required, we construct a correction layer, $C$. This model $C$ is designed to function as an efficient, specialized adaptation layer such that $M_1$ and $C$ applied together on $X_2$ replicate (or at least approximate) the outcomes $y_2$. Figure~\ref{fig:correction_layer} illustrates this basic setup, positioning the correction layer C as the bridge between models $M_1$ and $M_2$.

The core utility of creating this intermediate model $C$ lies in its interpretability: by focusing $C$ on the residual difference between the pre-shift and post-shift models, we transform the correction layer into a powerful diagnostic tool. First, assuming an interpretable modeling approach, the differential model $C$ can yield transparent insight into the precise driving forces that contribute to the systematic differences between $M_1$ and $M_2$. Second, these insights, together with the model $C$, can be used for scenario analysis. The latter is in particular of interest in cases where a major structural event happened between the time $t_1$ and $t_2$. In the example that we are going to discuss in this article, this major event will be the financial crisis of 2008.

The structure of the article is as follows. We first present background on {\em probabilistic rule models} (PRMs) and how they were built in the context of our application. Then we turn our attention to the specific construction of the correction layer based on PRMs. A main advantage of our approach lies in the fact that even in situations where the originating models $M_1$ and $M_2$ are {\em black-box} models, the fact that we are building the correction based on PRMs (which are entirely interpretable) allows us to draw insights into details of the difference of the two models. In the following section we demonstrate explicitly how to construct a PRM-based correction layer using Fannie Mae housing data~\cite{FannieMae}. The basic idea of this example is to consider a model $M_1$ trained on data from 2006 and then a model $M_2$ trained on data 2010. Given the financial crisis from 2008, we expect these models to be significantly different. Based on this, we construct a correction layer $C$ such that the model $M_1$ together with the correction layer approximates the model $M_2$. We then analyze the performance of the correction layer and illustrate how to use the correction layer to gain insights into the changes that affected the creditworthiness of particular customer segments. A conclusion summarizes the main results of the paper.

\section{Probabilistic Rule Models}

\subsection{Markov Logic and Rules}

In this paper, we will base the construction of the correction layer, the intermediate model $C$ described above, using the framework of probabilistic rules. While, in principle, many different modeling approaches can be used for the construction of the correction layer, Probabilistic Rule Models (PRMs) have a great many advantages, particularly in environments requiring high transparency. PRMs are founded on
Markov Logic Networks (MLNs), which successfully combine the robustness of statistical learning (via Markov Random Fields) with the interpretability of logical rules. This architectural choice—grounded in Markov Logic~\cite{nilsson:1986,richardson-domingos:2006}, Markov random fields~\cite{pearl:1988,wang-komodakis-etal:2013,venugopal-gogate:2014,bach-broecheler-etal:2017}, and probabilistic graphical models~\cite{koller-friedman:2009} —enables PRMs to maintain strong predictive accuracy while offering intrinsic interpretability. This intrinsic quality, where the rules themselves constitute the explanation, is a key requirement for lenders in creditworthiness and finance. Unlike post-hoc explainability methods (e.g., LIME or SHAP), the PRM model provides direct, symbolic, and auditable logic, which is often considered indispensable for high-stakes decisions~\cite{rudin:2019} where regulatory validation is paramount. We note that PRMs have also been successfully applied in other areas where interpretability is indispensable, for instance in healthcare~\cite{dhaese-finomore-etal:2021}.

In a nutshell, PRM consists of a set of logical formulas (rules) $(R_1, R_2, \ldots, R_m)$ and corresponding weighting factors $(\psi_1,\ldots,\psi_m)$, defined on a set of binary variables $(x_1, x_2, \ldots, x_n)$. In our formulation, $0<\psi_i < 1$, though other equivalent formulations are possible.

PRMs model a joined probability distribution over $\{x_i\}$ and, hence, belong to the class of generative models. Propositional Markov Logic
defines the probability of an evidence vector $\bx$ as
\[
    P(\bx) = \frac{1}{Z} \,\prod_i \phi_i(\bx)
\]
where $\phi_i$ are called potential functions, defined as
\[
    \phi_i(\bx) =
    \begin{cases}
        \psi_i & \text{if } \bx \text{ satisfies } R_i, \\
        1-\psi_i & \text{if } \bx \text{ falsifies } R_i.
    \end{cases}
\]
Here $Z$ is the normalization factor:
\[
    Z = \sum_j \prod_i \phi_i(\bx^j)
\]
where the summation goes over all possible assignments of truth values.

Markov Logic framework can be applied for predictive analytics, such as credit risk or fraud predictions. In a typical machine learning task, all variables are split into a set of input (explanatory) variables $\{\be_i\}$ and one or more target variables $y$: $\bx = (\be, y)$. In credit risk example, the target variable can be a default on a loan, and explanatory variables may include credit score, credit history, debt-to-income ratio, etc. The goal of predictive analytics is predicting probability of the target variable given the evidence of explanatory variables $P(y=1|\be)$.

In lending, PRM usually consists of association rules of the form
\[
    R_i:\quad (e_{k_1} \wedge e_{k_2} \wedge \ldots )\, \Rightarrow \, y\quad \text{with weight } \psi_i
\]
To the set of association rules, we will be adding an ``intercept'' rule:
\[
    R_0: \,\,y \quad \text{with weight } \psi_0
\]
The intercept rule ensures the correct a-priori probability of the target variable (provided, the model is well-calibrated).

Most commonly, single-factor rules of the type $x\Rightarrow y$, or two-factor rules $(x_1 \wedge x_2)\Rightarrow y$ are used. We say that the rule is \emph{shallow} if it has one factor in the premise.

We say that an association rule \emph{triggers} if the l.h.s. (the premise of the logical implication) asserts true. Rules that do not trigger, also do not contribute to the probability of the target variable. Triggered rules with $\psi > 0.5$ increase the probability, whereas rules with $\psi < 0.5$ decrease the probability.

Often, in context of lending, it is common to use points notation instead of probabilistic weighting factors (for instance point-based scorecard approach~\cite{anderson:2021}). For the sake of convenience, below instead of weighting factors $\psi$ we will be using points $p$ defined as
\begin{align*}
    & p_i = -a\,\ln\left(\frac{\psi_i}{1 - \psi_i}\right)
\end{align*}
The negative sign is chosen such that points have a positive connotation. Positive points increase, whereas negative points decrease the repayment probability. Here $a$ is an arbitrary scaling factor. 

\begin{figure}[ht]
    \vskip 0.2in
    \begin{center}
        \centerline{\includegraphics[width=0.9\columnwidth]{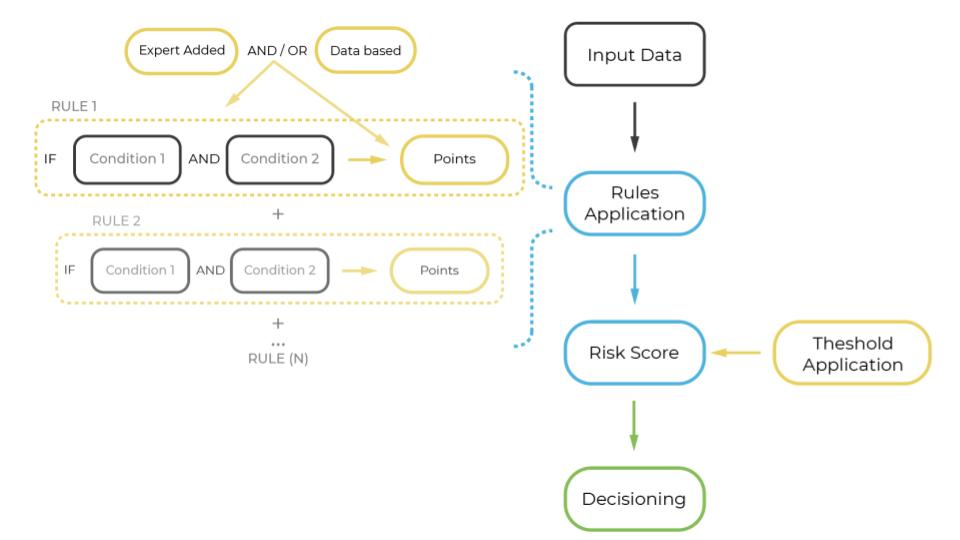}}
        \caption{Basic structure of a particular implementation of a workflow with probabilistic rule model. In the case considered here, all rules are found from analyzing the given input data although, in general, rules can also be added by domain experts.}
        \label{fig:PRM}
    \end{center}
    \vskip -0.2in
\end{figure}

The application of PRM in context of lending is illustrated schematically in figure~\ref{fig:PRM}. From the set of rules, using probabilistic logic, a score is computed. In particular, this score can be a probability of default (or repayment). Then, based on the chosen threshold, the lending decision can be made.

Let's consider an example rule in PRM:
\[
    \text{cscore } < 706 \qquad \Rightarrow \quad       -1.98 \,\text{ points}
\]
An intuitive interpretation of this rule is as follows: The rule indicates that if triggered (the variable ``cscore'' refers to the applicant's FICO score), i.e., a customer satisfies the conditions in the premise “FICO score below 706”, the creditworthiness score will change by a certain level that is determined during calibration of the model and this level is characterized by the number of points added or subtracted.

Notice the difference between probabilistic rules framework and a typical rigid rules based decisioning algorithm (usually implemented with a business rule engine -- BRE). A BRE-model consist of a set of the so-called knock-out logical rules: whenever a rule triggers, it demands the target variable to be 1, leading to an automatic rejection of a loan. In probabilistic framework, every rule contributes to the overall score, but every negative rule may be offset by other positive rules, leading to a much more accurate probability assessment, and more informed decision.

\subsection{Implementation and Calibration}

Building a PRM model includes the following steps:
\begin{enumerate}

    \item Data discretization. Since logic framework is only applicable to a binary data, all continuous variables have to be discretized. Sometimes binning of categorical variables is also required, especially for high cardinality variables. After discretization, one hot encoding can be used to convert the raw data into a binary format.

    The finer the discretization, the more accurate will be the PRM model. However, depending on the concrete application, it might be preferable to trade-off marginal accuracy increase in favour of model's simplicity and interpretability.

    \item Rule mining. There is a variety of available techniques to find a set of predictive rules. One can use association rule mining~\cite{agrawal-imielinski:1993,silverstein-brin-etal:1998,narvekar-syed:2015}, extract rules from a decision tree or construct rules manually based on domain expertize. This is where the approach demonstrates
    its great flexibility.

    \item Calibration. Once the rules are designed, the weighting factors (or points) have to be determined based on the historical training set, such that the model's output can be interpreted as a real-world probability of the target variable. A PRM model can also be used as a regressor. In this case the target variable in the training set can be continuous, and the model output is interpreted as an estimate of the continuous variable, as opposed to the probability of a binary target variable.

    Model calibration can be performed using gradient descent in combination with an appropriate choice of objective function. In this research we used L-BFGS algorithm~\cite{fletcher:1987} for the gradient descent implementation. For the classification mode, we use a regularized cross-entropy objective, and for the regression mode we use the regularized least squares objective:
    \[
        L = \sum (y_i - \hat y_i)^2 + \lambda \sum_{j=1}^m (\psi_j - 0.5)^2
    \]
    where $\hat y_i$ are model predictions, and $y_i$ are the ground truth values.

\end{enumerate}

We are now ready to discuss the particular PRM implementation as a correction layer, used in the application presented in this article.

\section{Building a Correction Layer with Probabilistic Rule Models}
\label{background}

Let us focus on mortgage default prediction models utilizing housing data. In a typical credit risk PRM model, the target variable, $y$, denotes loan default status: $y=1$ signifies ``default'' and $y=0$ signifies ``repayment.'' The input data, $X$, comprises features such as annual income, FICO credit score, home insurance characteristics, and geographical area. Hyper-parameter tuning is often necessary to optimize model performance. A typical PRM comprises 10 to 30 single- or two-factor association rules.

Constructing a correction layer using a PRM involves a similar approach, but with a different output variable. This layer aims to reconcile the discrepancies between models $M_1$ and $M_2$. Therefore, the output variable is not binary but represents the score difference between $M_1$ and $M_2$.

Here are all the necessary steps in detail:

\begin{enumerate}
    \item Build models $M_1$ and $M_2$. Model $M_1$ is created using data $(X_1, y_1)$, and $M_2$ is created using data $(X_2, y_2)$ where the different data sets correspond to different time windows. Datasets are split into train and test chunks for validation.
    \item Combine the data sets $(X_1, y_1)$ and $(X_2, y_2)$ to create the data set data $(X_c, y_c)$, which will be the basis for creating the correction layer. This step is essential in the context of Transfer Learning, as this mixed dataset is used for rule mining to mitigate the effect of population drift and capture the stable, generalized rule structure that holds across both the source and target domains.
    \item Use the data $(X_c, y_c)$ for rule mining. The rules $(R_0, R_1, \ldots , R_m)$ of this model form the basis of the correction layer. Then the model is calibrated on $(X_2, z)$, where the new output variable $z$ represents the score difference of $M_2$ and $M_1$ where the scoring of both models is done on the dataset $X_2$. The variable $z$ explicitly defines the residual error that the correction layer $C$ must learn, allowing $C$ to efficiently focus its learning capacity exclusively on the non-stationary components introduced by the concept drift, which is a key goal of this interpretable adaptation layer.
    \item To assess the effectiveness of the correction layer, apply $M_1$ to $X_2$, then apply the correction layer $C$ to $X_2$ and add the scores. This combination approximates the scores obtained by $M_2$ applied to $X_2$.
\end{enumerate}

\section{Pre- and Postcrisis Mortgage Data}
\label{mortgages}

\subsection{Building the Correction Layer}

In this section, we construct the correction layer for the Fannie Mae housing data for the years 2006 and 2010, and use the interpretable nature of the PRM model to gain insights into the impact of the 2008 financial crisis on different population groups.

The choice of the pre-crisis (2006) and post-crisis (2010) temporal split is deliberate. The 2008 financial crisis resulted from a structural boom-and-bust cycle characterized by loose credit standards and a complex, opaque securitization process, leading to an abrupt, system-wide shift in borrower risk profiles and market dynamics. This event represents an extreme test for concept drift mitigation, mirroring the dynamic, non-stationary environments that regulated entities like Fannie Mae must now manage, often requiring several risk model updates annually.

Using the pre-crisis data from the second quarter 2006 as input data $(X_1,y_1)$ we build a model $M_1$ to predict defaults. Then, we built the second model $M_2$ on post-crisis data, using the data from the second quarter of the year 2010. In both cases we chose to employ XGBoost, but we could have chosen any other modeling approach as well. After having both models in place, we trained a PRM $C$ by applying the approach outlined above. There is a lot of freedom and choice about how to set up the PRM in order to create a correction layer -- in particular in terms of the rule depth and the number of rules considered. In this case study, we chose to train a model with 15+1 rules and allowing only for single factors in each rule. The last rule R-16 is the intercept rule. The complete set of rules alongside the corresponding points is represented in the Appendix in table ~\ref{sample-table-2}.

\subsection{Effectiveness of the Correction Layer}

\begin{figure}[ht]
    \vskip 0.0in
    \begin{center}
        \centerline{\includegraphics[width=\columnwidth]{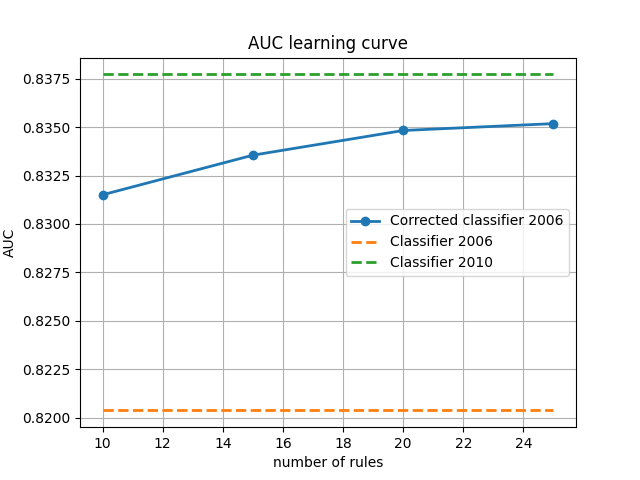}}
        \caption{Correction layer performance as a function of number of rules.}
        \label{fig:learning_curve}
    \end{center}
    \vskip -0.2in
\end{figure}

Performance of the correction layer model was tested on a holdout test set sampled from 2010 data. Measurement of AUC demonstrated a good convergence of the combination $M_1$ + correction layer to the $M_2$ model performance. In the figure~\ref{fig:learning_curve} we plotted the learning curve which shows how AUC depends on the number of rules in the correction layer. Further improvement can be achieved by refining the discretization and increasing the number and the depth of the rules.

However, we emphasize that achieving statistically significant equivalence with the fully retrained model ($M_2$) is not the primary objective of this architecture. For our goal—to diagnose the specific nature of the concept drift by extracting interpretable rules—a ``good enough'' performance that demonstrates strong convergence (as shown by the AUC and default rate alignment) is sufficient. The primary utility is the insight derived from the rules.

\begin{figure}[ht]
    \vskip 0.0in
    \begin{center}
        \centerline{\includegraphics[width=\columnwidth]{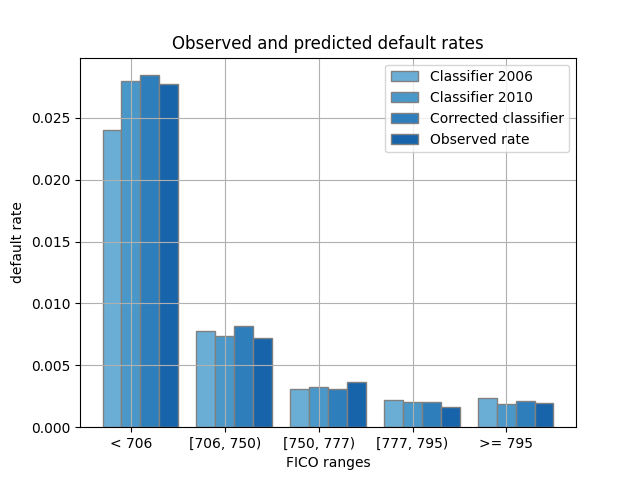}}
        \caption{Effectiveness of the correction layer: When the correction layer scores are added to the predictions of $M_1$,
            the combined score is very close to the predictions of $M_2$.}
        \label{fig:Correction_Layer_FICO}
    \end{center}
    \vskip -0.2in
\end{figure}

Another illustration of the correction layer performance is shown in figure~\ref{fig:Correction_Layer_FICO}. There we plotted the average default rate in 2010 data for 5 FICO bands, from low to high, alongside models predictions. We see that the 2006 model significantly underestimates the default rate for low-FICO band, but in combination with correction layer, it closely matches the observed probabilities.

\subsection{Insights from the Correction Layer}

Let's illustrate the diagnostic insights these rules can uncover by examining specific systematic changes captured by the correction layer $C$. The overall probability of default between 2006 and 2010 dropped significantly (from 2.88\% to 0.58\%). However, this massive reduction is primarily explained by population drift—banks introduced much stricter credit policies in an attempt to reduce the overall risk of their portfolios, removing high-risk applicants. Crucially, the changed economic reality also altered the underlying creditworthiness of people even with the same parameters. The correction layer is uniquely effective at isolating this structural shift.

For instance, let us examine the first rule R-01 (see Table~\ref{sample-table-2}):
\[
    \text{cscore } < 706 \qquad \Rightarrow \quad       -1.98 \,\text{ points}
\]

Here, the variable \emph{cscore} represents the average FICO credit score of the borrower and co-borrower. Negative points indicate decreasing probability of repayment (increased probability of default). The interpretation of these rules is straightforward, the results, however, might be surprising: A reduction of points means that the 2006 model overestimates the creditworthiness of a candidate in the low credit-score band if applied in 2010 reality.

The underpins the diagnostic value of the correction layer $C$, in this particular example focusing on the rule R-01: The result—that borrowers in the lower-middle FICO band (below 706) experienced a relative drop in creditworthiness—might appear counter-intuitive, but the PRM rule serves as an interpretable diagnosis of structural market failure. This finding is consistent with the results of other academic studies of the crisis, in particular the suggestion that the securitization process, which underpinned the housing boom, led to reduced screening effort primarily for loans with relatively high FICO scores (Prime/Alt-A) because these were easier to securitize and sell~\cite{ospina-uhlig:2018}. The underperformance of these ostensibly safer segments meant that post-crisis, the risk of borrowers just below the prime threshold (FICO $< 706$) increased relative to the new standard. The correction layer captures this specific shift: the $M_1$ model, which relied on pre-crisis screening norms, underestimated the true risk of this sub-population in the new regime. This insight is also confirmed by direct measurement of model predictions: the actual default rate of people with FICO $< 706$ in 2010 is 2.77\%, which is close to what the model $M_2$ predicts (2.79\%). The prediction of model $M_1$ for this group gives 2.4\% chances of default, which indicates that the model $M_1$ underestimates the riskiness of this group in the environment of the year 2010. Figure~\ref{fig:Correction_Layer_FICO} confirms this observation as well.

By looking at other rules of the correction layer, it is easy to see the overall impact of the financial crisis on the population. Each single-factor rule defines a cluster of population that triggers that rule, and a corresponding points attribution indicates what was the impact of the financial crisis on this population group. For instance, the rule R-13 shows that borrowers with relatively high debt-to-income ratio are also more likely to default in 2010 than the model $M_1$ would expect. On the other hand, the rule R-14 shows that if the loan to value (LTV) is relatively small, defaults are now less likely to occur in comparison with the predictions of the 2006 model. However, notice that LTV characterises the loan terms and not the customer. Therefore, this rule provides limited insight into the financial crisis' impact on the population.

In some cases, the insight provided by a rule could be discovered by a direct statistical measurement. In other situations, direct statistical discovery might be much more complicated because rules almost never trigger in isolation or due to population drift. Let's have a look again at the rule R-01 which describes the impact on the low FICO score band. We already know that this group of population has been negatively affected by the financial crisis. However, just by comparing the default rates in that group, we observe that it has dropped from 6.36\% in 2006 to 2.77\% in 2010, which could yield a misleading conclusion.

The reason for this potentially misleading result lies in the separation of drift effects. The observed drop in the raw default rate is the outcome of two countervailing forces: First, the stringent post-crisis underwriting policies (population drift) ensured that the remaining applicants in the low-FICO cluster possessed significantly better attributes than their pre-crisis counterparts, thereby reducing the raw default rate. Second, the fundamental deterioration of the economy (concept drift) inherently increased the true probability of default for any individual with FICO score below 706. The naive statistical comparison (6.36\% $\rightarrow$ 2.77\%) cannot isolate these two effects, failing to reveal what would happen to a low-FICO borrower if all other attributes were held constant. The interpretable correction layer resolves this diagnostic dilemma: The FICO rule (R-01) reveals the pure deterioration of creditworthiness, while the impact of all other stabilizing attributes is encoded in other specific rules, and the overall base rate change is reflected in the Intercept Rule (R-16).

To address this limitation and statistically isolate the effect of concept drift, analysts must examine much more specific clusters where the values of most attributes are ``fixed.'' This process attempts to compare ``exactly the same'' loan segments across the 2006 and 2010 environments. However, the drawback of this brute-force approach is significant: the number of required groups grows exponentially as more attributes are fixed, causing the number of cases falling into each group to drop exponentially. Consequently, the statistical analysis quickly becomes highly noisy or completely impossible due to data scarcity within these narrow pockets of the population.

How could one identify good pockets of population, which would be informative for the detailed portfolio analysis (such as in the current work, but not only)? The intrinsic interpretability of the PRM-based correction layer provides the direct answer to this question. By design, probabilistic rules automatically identify the most significant and structurally robust combinations of features that contribute to the systematic divergence between the two models.

\subsection{Rules clustering}

To better understand the factors that drove the impact of the financial crisis, it is instructive to examine smaller segments of the population that trigger specific combinations of rules, which we refer to as rules clusters. Each rules cluster naturally defines a corresponding cluster of the population, consisting of those individuals who trigger all the rules in question.

Often, banks are interested in groups of customers that are similar according to a specified metric. Typically, the similarity is defined in feature space. The PRM framework, in contrast, offers a powerful diagnostic lens by defining similarity based on the underlying, structural changes in credit risk captured by the correction layer: PRMs offer a different way to look at clients by considering customers similar if they trigger a similar set of rules. This idea can be combined with a cluster analysis of the rules
themselves in order to figure out first frequent combination of rules. If the rules fire frequently together, they are good candidates to form a cluster.

We can assess a quality of a cluster by comparing the default probability predicted by the cluster rules with the observed probability. If those probabilities closely match, this means that the rules defining the cluster are sufficient to predict the expected default rate in that group. If predicted and observed probabilities differ significantly, the cluster has to be refined by splitting it into sub-clusters, each triggering some additional rules.

\begin{figure}[ht]
    \vskip 0.0in
    \begin{center}
        \centerline{\includegraphics[scale = 0.5]{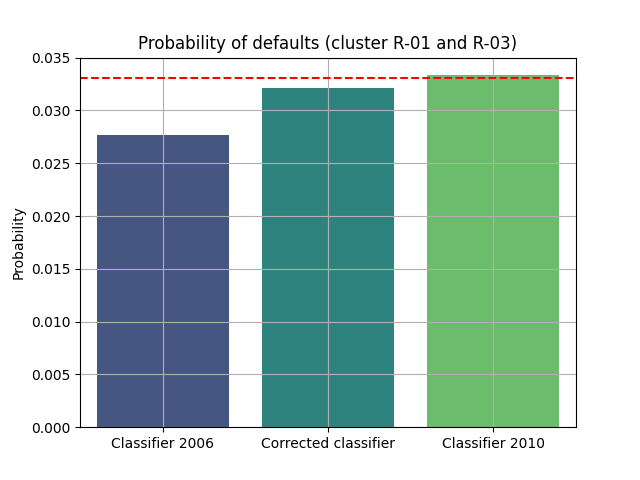}}
        \caption{Score comparison for the customer segment corresponding to the $(R_1,R_3)$-cluster, considering single borrowers ($R_3$) with a FICO Score of less than 706 ($R_1$).}
        \label{fig:Cluster_R1_R3}
    \end{center}
    \vskip -0.2in
\end{figure}

\begin{figure}[ht]
    \vskip 0.0in
    \begin{center}
        \centerline{\includegraphics[scale = 0.5]{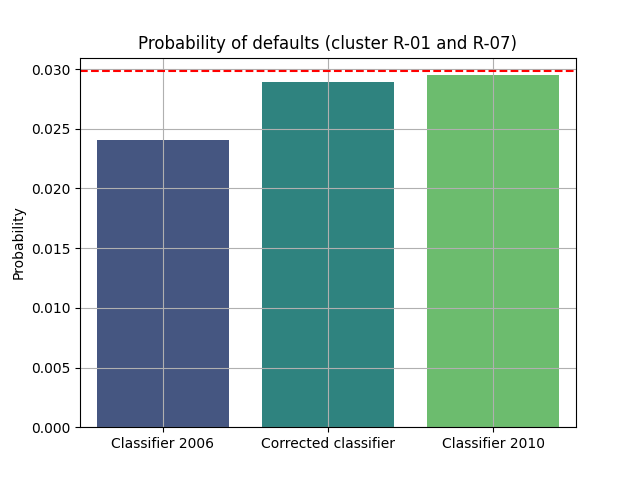}}
        \caption{Score comparison for the customer segment corresponding to the $(R_1,R_7)$-cluster, consisting of applicants with a FICO Score of less than 706 ($R_1$) and with a Single Family property type ($R_3$).}
        \label{fig:cluster_R1_R7}
    \end{center}
    \vskip -0.2in
\end{figure}

Examples of frequent rule combinations in our data set are (R-01, R-03) and (R-01, R-07). The first pair considers borrowers with a low FICO Score (rule R-01) who also have no co-borrowers (rule R-03).
The second rule concerns borrowers with a low FICO Score and specifying the property type as ``Single-Family''. The following graphs~\ref{fig:Cluster_R1_R3} and~\ref{fig:cluster_R1_R7} show that the correction layer allows to predict the scores for these two customer segments. The statistical picture is also in-line with the expected effect of these two clusters, since all three rules R-01, R-03 and R-06 have negative associated points, which means a drop in creditworthiness.

\subsection{Preparing for the Future: Scenario Analysis}
\label{scenario}

Identifying the client base for which the changes due to the crisis were most prominent is only one application of the insights provided by the analysis of the data based on the correction layer. This process directly illustrates the benefits of our interpretably designed Transfer Learning approach. The fact that a combination of the pre-crisis model $M_1$ and the correction layer $C$ approximate the post-crisis model $M_2$ also can be used for scenario analysis: In a future scenario where a similar crisis is anticipated, the combination of a current model and the correction layer can mimic this future scenario which allows, for instance, for a detailed risk analysis for the case that the crisis should happen.

This scenario analysis capability is significantly enhanced by the intrinsic interpretability of the PRM correction layer. It allows risk managers to utilize the derived rules (such as R-01 and R-13) to perform informed simulations. Rather than relying solely on a simple correlation-based model, experts can adjust the weights or premises of the correction rules based on their expectations for a future economic regime—for example, anticipating whether new regulations will intensify the effects on specific DTI brackets (R-13) or alter the screening standards for Prime/Alt-A borrowers (R-01). Moreover, this scenario analysis can be overlaid with a predictive model of the likelihood of the crisis (e.g., a recession) to happen. Such an analysis is common when performing a stress test of a portfolio to evaluate its performance if a recession might occur in the future. Furthermore, the rule-based structure allows for the direct simulation of macroeconomic factor changes. For instance, experts can modify the points associated with rules related to loan purpose or credit score (R-01) to simulate the precise impact of rising unemployment rates, or adjust interest-rate related rules (R-06) to model the systemic effects of sharp central bank rate hikes. This fine-tuned, factor-specific approach drastically improves the utility of stress testing.

\section{Conclusion}

In this article, we presented an example of how to build a correction layer based on probabilistic rule models (PRMs) to analyze the difference of mortgage data from 2006 and mortgage data from 2010. The interpretable nature of the correction layer enabled us to gain detailed, sometimes counter-intuitive insights about several customer segments which help to assess portfolio risks more accurately. Crucially, this insight is able to isolate the impact of population drift (changes in the applicant pool due to stricter underwriting) from the impact on creditworthiness (true changes in riskiness, or concept drift), which would be otherwise extremely difficult to do with brute-force statistical analysis.

We established the property of the correction layer $C$ given by $M_1 +C \approx M_2$ as a vital design element of the strategy leveraging Transfer Learning principles for efficient and transparent model adaptation. The use of PRMs guarantees intrinsic interpretability, moving the correction layer beyond mere predictive convergence and transforming it into a diagnostic tool. By isolating the systematic shift between the pre-crisis and post-crisis lending regimes, the rules extracted by the correction layer successfully explained specific changes in borrower riskiness, which were shown to be consistent with external academic findings on the structural impact of securitization and reduced screening standards. This capability—providing auditable, symbolic explanations for concept drift—is paramount for maintaining robust risk management and regulatory compliance in high-stakes financial applications.

\section*{Appendix}

The following table lists the complete set of rules of the correction layer for the Fannie Mae housing data:

\begin{table}[htb]
    \caption{Complete set of rules of the correction layer for the Fannie Mae housing data}
    \label{sample-table-2}
    \vskip 0.1in
    \begin{center}
        \begin{small}
            \begin{sc}
                \begin{tabular}{lcr}
                    \toprule
                    Rule Identifier & Rule Definition           & Points \\
                    \midrule
                    R-01            & cscore $< 706$            & -1.98  \\
                    R-02            & orig\_rate $\geq 6$       & -0.97  \\
                    R-03            & num\_bo $< 2$             & -0.11  \\
                    R-04            & loan\_term $\geq 360$     & 0.47   \\
                    R-05            & purpose in [``U'', ``P''] & 0.7    \\
                    R-06            & orig\_rate in [5.25, 6)   & -2.92  \\
                    R-07            & prop\_type = ``SF''       & -1.05  \\
                    R-08            & purpose = ``C''           & -0.18  \\
                    R-09            & insurance\_pct $\geq 9$   & 1.50   \\
                    R-10            & comb\_ltv $\geq 80$       & -0.91  \\
                    R-11            & state = ``area1''         & 0.26   \\
                    R-12            & occupancy\_type = ``P''   & -0.4   \\
                    R-13            & dti $\geq 43$             & -0.87  \\
                    R-14            & comb\_ltv $< 55$          & 0.33   \\
                    R-15            & ltv in [78, 80)           & -0.76  \\
                    R-16            & -                         & 1.71   \\
                    \bottomrule
                \end{tabular}
            \end{sc}
        \end{small}
    \end{center}
    \vskip -0.1in
\end{table}

Remark 1: The rules in this table are sorted by the rule's impact, which is a combination of the associated points and a coverage, where the latter is defined as a percentage of the population that triggers the rule. In this table, the last rule R-16 is the intercept rule. In indicates the so-called basepoints, responsible for making the a-priori probability of the target variable match the observed default rate.

Remark 2: The data used for building the models, as well as the full description of variables is freely available on Fannie Mae portal~\cite{FannieMae}. For the sake of the reader's convenience, and since we shortened some variable names, below is a brief description of variables whose meaning might not be obvious:
\begin{itemize}
    \item \textbf{cscore}: an average credit score of the borrower and co-borrower.
    \item \textbf{orig\_rate}: an interest rate at origination.
    \item \textbf{purpose}: Cash-Out Refinance = C, Refinance = R, Purchase = P, \\ Refinance-Not Specified = U.
    \item \textbf{occupancy\_type}: Principal = P; Second = S; Investor = I; Unknown = U.
    \item \textbf{prop\_type}: condominium = CO; co-operative = CP; Planned Urban Development = PU; manufactured home = MH; single-family home = SF\@.
    \item \textbf{num\_bo}: Number of Borrowers.
    \item \textbf{state}: property state. In this work we grouped the states into 4 large categories. In particular, area1 = [``NV'', ``AZ'', ``CA'', ``FL'', ``MI''].

\end{itemize}

\bibliography{correction_layerbib}

\end{document}